# Is happiness u-shaped in age everywhere?
# A methodological reconsideration for Europe

Forthcoming in *National Economic Institute Review*

David Bartram

University of Leicester
Leicester LE1 7RH
United Kingdom
ORCID: 0000-0002-7278-2270

d.bartram@le.ac.uk

**Abstract:**

A recent contribution to research on age and well-being (Blanchflower 2021) asserts that the impact of age on happiness is "u-shaped" virtually everywhere. I evaluate that finding for European countries, considering whether it is robust to alternative methodological approaches. The analysis excludes control variables that are affected by age (noting that those variable are not antecedents of age) and explores the relationship via models that do not impose a quadratic functional form. The paper shows that these alternate approaches do not show a u-shape "everywhere": u-shapes are evident for some countries, but for others the pattern is quite different.

Declarations
- Funding: none
- Conflicts of interest: none
- Availability of data: data from the European Social Survey are available at www.europeansocialsurvey.org
- Code availability: the R syntax/code for this paper is available on request and at the following link: https://osf.io/3axq6/?view_only=0c55ece4430148f8a77a2522975fda79

Acknowledgements:

I am very grateful to David Ekstam for insightful comments on an earlier draft.

# 1 Introduction

Does happiness generally fall in middle age and then rise as people get older, in line with the popular 'u-shape' idea? A recent contribution to research on this question (Blanchflower 2021) offered as its core finding the idea that the relationship between age and happiness is indeed u-shaped virtually 'everywhere', including 45 of 46 countries in Europe.

More broadly, the u-shape idea is a matter of significant dispute. In some contributions the relationship is perceived as flat (Kassenboehmer and Haisken-DeNew 2012, Easterlin 2006). Some researchers find that happiness does not fall in early adulthood but rather rises (Galambos et al. 2015). Others find that happiness falls (instead of rising) in older age, especially as people become *very* old (e.g. Frijters and Beatton 2012, Gerstorf et al. 2008). Laaksonen (2018) finds that the pattern is typically more complex than a u-shape. The idea that there is a uniform pattern at country level is evaluated (and dismissed) by Bittman (2021), Galambos et al. (2020), and Bartram (2021a). In the context of these competing conclusions, a paper that finds a u-shape *everywhere* is intriguing, to say the least.

In this paper I reconsider that finding by analysing data on European countries using a set of methodological decisions that depart from those of Blanchflower (2021). Blanchflower's 'u-shapes everywhere' finding comes from models that restrict the analysis to people under the age of 70, impose a quadratic function for the age effects, and include control variables for individual characteristics/circumstances (sex, marital status, education, and labor force status). I focus first on the decisions to include the specified controls and to restrict the analysis to people younger than 70; I show that these decisions inflate the coefficients for the quadratic age function, pushing them away from zero and thus increasing the impression of a u-shape. In line with Glenn (2009), Hellevik (2017), and Bartram (2021a), I argue that inclusion of controls for individual circumstances induces bias in the estimate of a causal impact of age on happiness (age→happiness is a useful shorthand); there are good reasons (presented below) to prefer models without these controls. For analysis of European countries I also argue for not restricting the age to a maximum of 70, given that life expectancy in Europe is generally rather higher than that.

I then present models that portray the age→happiness relationship via age ranges (categories), rather than using the conventional quadratic functional form. Use of a specified functional form imposes an unnecessary constraint (Bittman 2021, Kratz and Brüderl 2021). If the actual relationship is in line with the specified function, then that relationship will also show up in an analysis that does not impose the constraint. If, however, the actual relationship is *not* consistent with the specified function, then the constraint is likely to act as a distortion, an impediment to discerning the underlying relationship. Compared to an analysis that is less restrictive in functional form, a specific function is likely to be all pain and no gain, certainly if the actual relationship is not consistent with that function. Using age ranges, it is also more feasible to include a control for cohort, as a way of ensuring that variation assigned to an age variable is not in fact reflecting differences rooted in people's early-life experiences.

The main focus of this article is methodological, so I do not review previous findings in substantive/theoretical terms (i.e., reasons we might expect to find u-shapes, or not find them). To keep things relatively simple, I focus only on Europe, use a single dataset (the European Social Survey, rounds 1-8), and analyse only one dependent variable (happiness). If readers find the methodological arguments convincing, the approach used here can (and surely should) be applied to other regions, datasets, and dependent variables.

# 2 A methodological evaluation and revision

In a very general sense, there is no single 'right' way to analyse data for any specific research question. All methodological decisions come with advantages and disadvantages. To foster confidence in our decisions, it helps to present readers with the advantages and disadvantages explicitly, and to demonstrate the consequences of our decisions via comparison to results reached via different decisions. In this section I explore two key methodological issues noted above (control variables and restriction of age range), first via arguments and then via a comparative analysis of results for a single country (Germany). Having reached a detailed conclusion about the 'best' way to analyse data for estimating age→happiness (not a *perfect* way, but better than the alternatives), I then proceed in the following section to present a more comprehensive set of results for European countries.

## *2.1 Control variables*

In general, quantitative researchers would almost always use control variables in analyses that consider the relationship between one variable (X) and another (Y), especially if the relationship is to be interpreted as causal (X→Y). Unfortunately, a great deal is typically taken for granted about what control variables do (in regression models), and about how to select them (Bartram 2021b). In some quantitative contributions, no particular criterion for selection of control variables is articulated. Sometimes researchers are simply guided by precedent, using the same set of controls used by one (or a few) previous researcher(s). In recent years we see more attention paid to those topics (e.g. Pearl and Mackenzie 2018, Gangl 2010). It is possible (and surely necessary) to go beyond what we might call a 'conventional' perspective about controls.

If a criterion of any sort *is* articulated, that criterion is conventionally: control for 'other determinants' of the dependent variable (Y). That criterion is ineffective and potentially misleading. Having identified 'other determinants' of Y, it is then necessary to consider the relationship between the potential controls (here labelled W) and the focal *independent* variable (X). It bears emphasising that this is not (yet) the standard practice in the social sciences – though it is well established in epidemiology (e.g. Schisterman et al. 2009).

To discern the importance of taking this additional step we can consider a straightforward distinction between two possibilities:

1. Is the potential control (W) an *antecedent* of the focal independent variable? Is the pattern W→X?
2. Or, is the potential control *influenced by* the focal independent variable? In other words, is it X→W?

Control variables do not do just one thing. A great deal depends on this distinction. To see why, we can articulate what the purpose of statistical control is. The purpose is to avoid bias in our estimates of X→Y. The relationship we see in a bivariate analysis might reflect the influence of some third variable on X and Y. In the classic example, children's academic abilities are correlated with their shoe size – but it is a nonsense to conclude that one causes the other. Once we control for age, the correlation disappears; age is the true 'cause' of both. The example works because the control (age) is an antecedent of the other two variables; in particular, it is an antecedent of X (W→X), whichever variable is identified here as X (shoe size or academic abilities). The control works as intended to redress bias on this basis.

However, when the potential control is *influenced by* the focal independent variable (X→W), the control does not work to redress bias; instead, the control *exacerbates* bias. Suppose we want to estimate unemployment→happiness. One of the 'other determinants' of happiness is income. Should we control for income? Income is not an important antecedent of unemployment (our X here); people can get sacked from any sort of job, regardless of salary. The relationship is better captured by X→W: when someone gets sacked, their income goes down (often dramatically), and the lower income leads to lower happiness. If we control for income when estimating unemployment→happiness, we compare people who are unemployed to people who are employed while holding income constant – i.e., the comparison takes place among people earning the same level of income. The estimate of unemployment→happiness now obscures part of the actual effect of unemployment on happiness, the portion that travels via unemployment's impact on income. We likely need other controls, but it is an error to include income as a control here.

So, to estimate X→Y, we need controls where W→X, and we must *exclude* controls where X→W (Bartram 2021b). (Perhaps some potential controls are unrelated to X. In that case, they are irrelevant, even if related to Y; they are not needed to estimate X→Y, though including them will not harm the estimate.) A consideration of how W relates to X is essential; it is not enough to consider how W relates to Y.

So, to identify the controls needed to estimate age→happiness, we need to ask: what are the antecedents of age? The only sensible answer is: there are none. Until they die, *everyone* keeps getting older, at exactly the same rate, no matter what their other characteristics. None of the 'other determinants' of happiness affect how old someone is, or the rate at which they get older (in the usual numerical sense). Some characteristics or situations might have an impact on mortality/life-span – but for people who are still alive (and are thus available to participate in surveys) those variables have no impact at all on their age. There are no controls for which W→X.

What's worse, we are decidedly in the realm of X→W. Ageing has an impact on a wide range of other aspects of people's situations. A particularly relevant one here is marital status. As people get older, they are more likely to experience the death of a spouse/partner, an event with negative consequences for their happiness. If we control for marital status when estimating age→happiness, we will obscure a portion of age's impact on happiness, the portion that travels via age's impact on marital status. The situation is analogous to the example above (unemployment→income→happiness).

For estimation of age→happiness, we are in a situation that is open to misunderstandings. No controls are needed to estimate that effect. (We still need a discussion of cohort and period; see below.) Blanchflower (2021) and Blanchflower and Oswald (2009), responding to Glenn's (2009) critique of Blanchflower and Oswald (2008) on this point, say that a model with no controls is (merely?) a 'descriptive' analysis. They argue for use of controls to achieve a 'ceteris paribus analytical' finding. This rhetoric leads us astray in this context. Instead, what we see here is best understood as a special case. For most causal estimations, it does make sense to include some control variables in one's model. But when there are no antecedents of X, no controls are needed for an estimate of X→Y (that's what makes this case 'special'). A specification of this sort does not make a model 'descriptive'; it makes it *correct* – especially relative to a model that includes controls where X→W. To make sense of the idea of 'ceteris paribus', we need to ask: *which* other variables are being held constant, and *why*? Inclusion of inappropriate control variables does not yield a 'ceteris paribus analytical' model that is superior to a model containing no control variables, in situations where there are no antecedents of X.

From a different angle, an analysis that uses controls is not incorrect; it simply offers different information. If we include a control where X→W, we get a result for X→Y that can be interpreted as a 'direct effect'. Some might say that an analysis of age→happiness that includes controls gives us a result reflecting a 'pure' impact of age. But we should have clarity on what a result of that sort has been 'purified' *of*. What we can see from the discussion above of marital status is that a direct effect is not net of the impact of 'other' variables – it is net of the part of the effect *of age itself*. Results from an analysis that creates direct effects should be carefully articulated in those terms; it is not the same as a 'total' effect – so, it is not *the effect* of age.

I demonstrate below the consequences of using inappropriate controls – but I turn first to a discussion of whether it makes sense to restrict the analysis to respondents younger than 70.

## *2.2 An age restriction?*

Blanchflower (2021) offers a global analysis, covering 145 countries. Life expectancy is of course lower in some countries, relative to life expectancy typical in very wealthy countries. To ensure consistency of results across a broad range of countries (and in consideration of the small sample sizes available for older people in some countries), Blanchflower restricts his analysis to respondents younger than 70.

This methodological decision is by no means 'incorrect'. It is however consequential. We can draw on existing research to predict a likely consequence. As against a 'u-shape' pattern, some analysts (e.g. Brockman 2010, Beja 2018, Frijters and Beatton 2012, Gerstorf et al. 2008) find an 's-shape' pattern: happiness is higher among younger people, declines towards middle age, rises as people get older, but then declines again as people get very old and start to experience significant challenges (declining health, widowhood, etc. – see e.g. Hudomiet et al. 2020). An analysis that includes not just an age term and an age-squared term but also an age-cubed term reveals that pattern in some instances.

If (or where) that pattern prevails, then restricting the upper age bound to 70 is likely to increase the impression of a u-shape, relative to an analysis that does not impose that restriction. The argument *for* the restriction is sensible, in the context of an analysis that covers a very broad range of countries. But we should be mindful of the impact the decision will have for countries where people live for substantial periods past the age of 70. The decision is not 'wrong' in general, but if we want to know whether (and to what extent) age→happiness is u-shaped in wealthy countries we likely have reason to prefer an analysis that does not impose an upper age restriction.

## *2.3 Demonstration*

I now show the consequences of these two methodological decisions in Blanchflower (2021), via an analysis that uses the same (quadratic) functional form. I focus for now on one country: Germany (selected simply as a typical example, to demonstrate patterns that are also readily apparent in other countries). I draw on data from the European Social Survey (rounds 1 through 8, corresponding to 2002 to 2016; see Jowell 2007). The survey mode is face-to-face, using random probability selection methods at each stage of the multi-stage design; average response rate for the participating countries in the most recent round is 55.4% (rates are taken directly from the survey project website). Happiness is drawn from a question asking 'Taking all things together, how happy would you say you are?', with 11 available response options (0 through 10). Age is given in years; including an age-squared term as well gives us the usual functional form. All models include a period variable ('round' from the ESS). For the model that includes control variables, we have sex (male vs. female), education (five categories), marital status (six categories including 'other'), and labor force status ('main activity', eight categories, with community/military service condensed into 'other' on grounds of very small numbers). All models include the design weights offered with the dataset.

Table 1 starts by presenting Blanchflower's own estimate for Germany, drawn from a model (1) that includes the indicated controls. I present my own version of that model (2) as a replication; the coefficients are very close to Blanchflower's. The third model removes the indicated controls (keeping the period variable). The first three columns entail the restriction of age<70; the fourth column removes that restriction.

**Table 1: Models of happiness (Germany)**

|  | Model 1 (Blanchflower) | Model 2 | Model 3 | Model 4 |
|---|---|---|---|---|
| Age | -0.11600 | -0.11463 | -0.04384 | -0.02073 |
| T | 15.26 | 15.35 | 8.73 | 6.10 |
| Age-squared | 0.00119 | 0.00120 | 0.00047 | 0.00017 |
| T | 14.00 | 14.29 | 7.93 | 5.07 |
| Constant |  | 9.437 | 8.182 | 7.789 |
| Controls? | Yes | Yes | No | No |
| Includes age 70+? | No | No | No | Yes |

Note: all models contain a control for period (survey year).

Comparing Model 2 to Model 3, we see that removal of the controls cuts the age and age-squared coefficients in half (a pattern very much in line with results in Frijters and Beatton 2012). On the basis that the pattern describing the controls is X→W, we can say that the controls induce significant bias in the estimation of age→happiness (as given in Model 2).[1]

Comparing Model 3 to Model 4, we see that removal of the age<70 restriction has a further substantial impact on the age and age-squared coefficients: they are cut in half again. The word 'bias' is perhaps less suitable as a way of characterising the differences between the two sets of results. But given that many people in Germany live rather longer than 70 years, it seems more sensible to model the age→happiness relationship via an analysis that considers the full age span. In that analysis, the extent of u-shape is smaller (with coefficients closer to zero). Via both adjustments, the coefficient for age has been reduced by 80 per cent, and the coefficient for age-squared has been reduced by 79 per cent.

To appreciate further the consequences for 'u-shape', in Figure 1 I plot the curves from the different model results. The dashed line comes from the model containing control variables and restricted to age<70 (corresponding to Blanchflower's analysis); it shows the 'deepest' u-shape. The dotted line, from the model excluding controls, is shallower. The solid line, with the age restriction removed, is shallower still.

To characterise age→happiness for Germans, in my view the model without control variables and without the age restriction is preferable to the models with controls and/or with an age restriction, for the reasons given above. It is of course not a 'perfect' model; in fact I will shortly argue that it has certain disadvantages and we should do something different. But the 'something different' builds on the perspective that says Model 4 is better than Models 1, 2, and 3.

---

[1] A control for sex is irrelevant; when sex is added as a control to Model 3, the coefficients for age and age-squared are virtually identical. This is not where the bias is coming from.

**Figure 1: age-happiness curves (Germany) from three quadratic models**

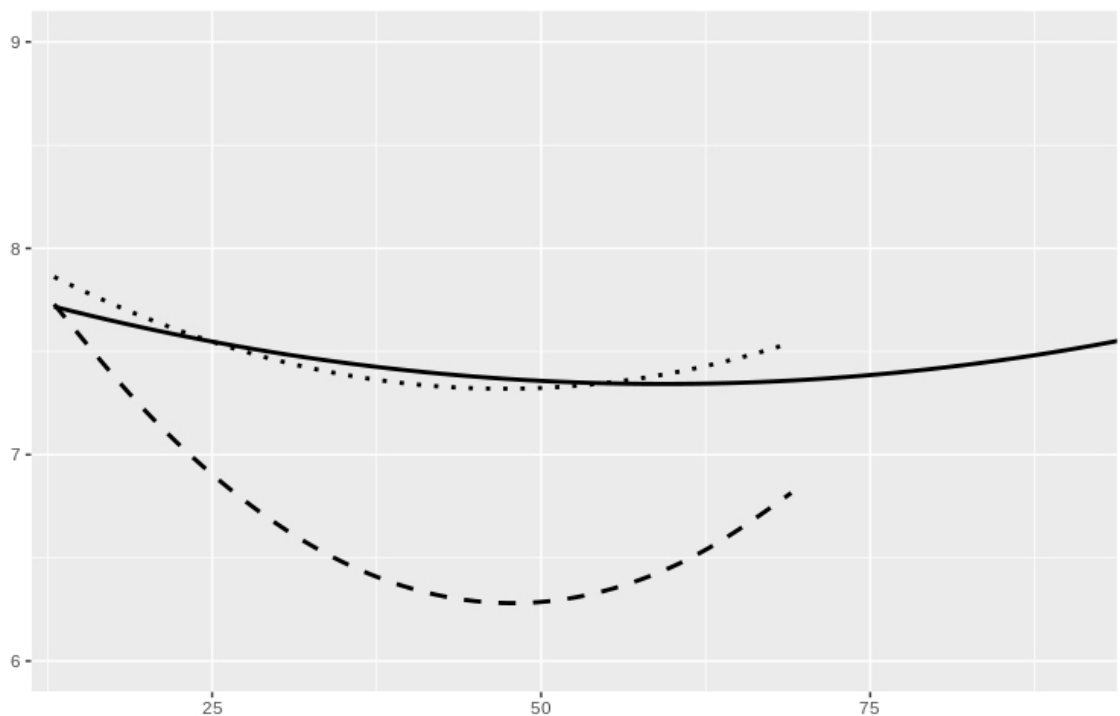

Table 2 presents models equivalent to Model 4 above (for Germany) for almost all the European countries for which ESS data are available. (For Croatia and Luxembourg the available data are limited, coming from only two adjacent rounds; this time-range does not give good leverage in connection with the need to control for period and cohort effects.) In Blanchflower's analysis of ESS data, virtually all of the age and age-squared coefficients come with t-statistics larger than 1.5, the threshold he identifies for statistical significance (the only exception was the age coefficient for Poland). In Table 2, countries for which t-statistics are greater than 1.5 (for both variables) are shaded, to indicate instances where a u-shape can still be identified via this threshold (in models that exclude controls and include people older than 70). For 23 countries, a conclusion reached via statistical significance leads to identification of a u-shape. For 7 of the 30, however, the finding of u-shape is no longer supported. This is our first indication that the u-shape is not found 'everywhere'.

Where u-shapes *are* evident, in each instance the curve is shallower, indicating a smaller reduction of happiness in middle age and a smaller rise afterwards. The final two columns show the per cent reduction in size of coefficient, relative to models (not shown) that include controls (where X→W) and impose the restriction of age<70. On average, the reduction for age is 66.4 per cent, and for age-squared 77.2 per cent. In a few instances (Austria, Israel, and Italy) the per-cent reduction for change in the age-squared coefficient is more than 100 per cent, which indicates that the sign of the coefficient has changed. For those three countries the coefficient for age-squared is now negative – indicating that happiness in those countries does not rise after middle age but instead continues to decline (though only for Italy is the age-squared coefficient statistically significant). For Denmark, in contrast, the sign of the age coefficient changes from negative to positive; in that country happiness appears to increase across the entire life-course (another form of departure from u-shape), though here as well not 'significantly'.

**Table 2: OLS estimates of happiness (excluding controls, all ages)**

|  | Age | T | Age-squared | T | % reduction in magnitude of coefficients, relative to models with controls and age<70 | |
|---|---|---|---|---|---|---|
|  |  |  |  |  | Age | Age-squared |
| Austria | -0.00540 | 1.05 | -0.000004 | 0.08 | 90.4 | 100.8 |
| Belgium | -0.01441 | 4.20 | 0.00012 | 3.49 | 75.7 | 81.3 |
| Bulgaria | -0.10101 | 11.72 | 0.00061 | 7.06 | 48.5 | 63.0 |
| Switzerland | -0.00923 | 2.72 | 0.00010 | 2.83 | 84.0 | 83.6 |
| Cyprus | -0.01903 | 2.37 | 0.00007 | 0.82 | 68.5 | 88.3 |
| Czech Rep. | -0.04085 | 8.73 | 0.00024 | 4.87 | 35.0 | 53.8 |
| Germany | -0.02073 | 6.10 | 0.00017 | 5.07 | 81.9 | 85.8 |
| Denmark | 0.00027 | 0.07 | 0.00005 | 1.20 | 100.5 | 92.4 |
| Estonia | -0.04610 | 9.47 | 0.00023 | 4.70 | 43.2 | 62.3 |
| Spain | -0.02925 | 7.68 | 0.00016 | 4.17 | 69.8 | 81.8 |
| Finland | -0.00807 | 2.52 | 0.00004 | 1.21 | 87.4 | 93.4 |
| France | -0.03857 | 9.08 | 0.00028 | 6.56 | 55.1 | 63.6 |
| Britain | -0.03487 | 9.11 | 0.00042 | 10.89 | 60.6 | 58.8 |
| Greece | -0.03767 | 6.13 | 0.00019 | 3.08 | 68.7 | 82.9 |
| Hungary | -0.06527 | 11.42 | 0.00040 | 6.89 | 41.8 | 58.3 |
| Ireland | -0.03118 | 7.96 | 0.00039 | 9.47 | 66.9 | 63.9 |
| Israel | -0.01130 | 2.60 | -0.00002 | 0.34 | 87.7 | 102.3 |
| Iceland | -0.00184 | 0.21 | 0.00008 | 0.92 | 96.5 | 86.7 |
| Italy | -0.00214 | 0.29 | -0.00014 | 1.86 | 97.2 | 119.4 |
| Lithuania | -0.05822 | 9.41 | 0.00025 | 3.85 | 22.1 | 50.0 |
| Netherlands | -0.00919 | 3.03 | 0.00007 | 2.42 | 84.5 | 88.9 |
| Norway | -0.01343 | 3.71 | 0.00014 | 3.87 | 79.6 | 79.4 |
| Poland | -0.04718 | 9.88 | 0.00026 | 5.25 | 53.9 | 71.1 |
| Portugal | -0.04111 | 9.39 | 0.00017 | 3.95 | 44.9 | 71.7 |
| Russia | -0.05210 | 9.20 | 0.00032 | 5.26 | 46.9 | 62.4 |
| Sweden | -0.00565 | 1.59 | 0.00007 | 1.95 | 91.1 | 89.2 |
| Slovenia | -0.04755 | 9.22 | 0.00024 | 4.57 | 50.3 | 68.0 |
| Slovakia | -0.05201 | 8.18 | 0.00035 | 5.28 | 44.8 | 60.7 |
| Turkey | -0.05163 | 4.30 | 0.00056 | 4.15 | 57.0 | 60.0 |
| Ukraine | -0.04766 | 6.90 | 0.00013 | 1.82 | 45.2 | 79.0 |
| Average |  |  |  |  | 66.4 | 77.2 |

Note: the models contain a control for period (survey year). Shaded countries are those where a u-shape is evident.

## 2.4 Beyond the quadratic functional form

An estimate with a defined functional form (e.g. linear, or quadratic) can be useful – *if* the underlying relationship is effectively captured via the function. A function is then a useful simplification, a way of conveying information in a single number (or, perhaps, two numbers, as with a quadratic function). With a small number of quantities to evaluate, it is then also more straightforward to draw conclusions via hypothesis tests (i.e., p<0.05).

Whether such an estimate is consistent with the underlying relationship should be checked, via comparison to an estimate that does not impose the function (compare Bittman 2021). (Simonsohn 2018 offers specific cautions against use of a quadratic function to detect u-shapes.) A model specifying a functional form might fit the data poorly but still yield results where p<0.05. A risk of that sort can be exacerbated simply by virtue of having a large sample. These points merit emphasis: with a large sample it is easier to get results where p<0.05, even when the specified function does not accurately portray the actual pattern in the data.

Blanchflower (2021) recognises the point, offering (in his Figure 7) an evaluation of age→happiness (with ESS data) that uses a series of age dummy variables instead of the quadratic function. The figure (which includes respondents older than 70) presents a distinct u-shape, with average happiness declining from 8.55 at age 18 to 7.60 at age 53 and then rising above 8 as respondents move into their 80s. It appears to offer strong confirmation of a u-shape for Europe.

There are two important observations to make about that figure. First: it comes from a model that includes control variables (so, the results are biased). And second, it covers all 32 participating countries together. It thus tells us that age→happiness is u-shaped for Europe as a whole (though in part that conclusion comes via use of controls where X→W), but it does not tell us that age→happiness is u-shaped for *all* of those countries (taken separately).

I now proceed to a country-level evaluation of age→happiness that dispenses with a quadratic function and uses age ranges. An analysis using age ranges is useful for an additional reason, beyond evaluation of the findings from quadratic models. Any model of an 'ageing' effect has to consider the challenge that arises from the confounding of age (A) with period (P) and cohort (C). An 'APC problem' is rooted in the fact that any one of those terms forms a perfect linear combination of the other two (A + C = P, etc.), when each is specified in year units (see de Ree and Alessie 2011 for a cogent discussion in the context of age and subjective well-being).

To put the point in more substantive terms: when we have a coefficient for age from a model that does not include terms for period and cohort, we would have to consider alternative interpretations. Does the age coefficient partly capture changes that are taking place over a certain period in time, for all, perhaps because of specific events with broad impacts (e.g. a pandemic or financial crisis)? Does the age coefficient partly reflect the fact that older (vs. younger) people were born during a certain era and thus had different formative experiences? To identify a distinct age effect, we need to disentangle age from period and cohort. *These* are the needed control variables.

Attempts to address the APC problem have produced a wide range of disparate approaches, each attempting to 'break' the linearity of the APC combination while controlling effectively for cohort and period. For a while, a 'hierarchical age-period-cohort' (HAPC) approach using a cross-classified random-effects model (CCREM) seemed promising (Yang 2008, Yang and Land 2008). In this approach, terms for period (in years) and cohort (in birth-year ranges) are entered as random effects, which helps enable model identification.

In recent years, however, there is growing awareness of some potentially important technical drawbacks (O'Brien 2017). The model can 'shrink' the variation assigned to one of the random effects (typically cohort), inflating the apparent age effect (Bell and Jones 2018, Luo and Hodges 2020). This technique is not a 'silver bullet'. There likely *is* no silver bullet; instead, we are again enjoined to construct models using alternate approaches and then evaluate them substantively, perhaps using 'side information' (Ekstam 2021).

Here is where we see another advantage of using a model that includes age ranges instead of a quadratic function. In this approach we can enter cohort and period also as fixed effects, foregoing random-effects (Ekstam 2021) and thus avoiding the 'shrinkage' problem. While still acknowledging that no approach is 'ideal' on technical grounds, a model along these lines is arguably the most conservative in a technical sense, avoiding highly complex algorithms for calculation that are still not well understood.

In Table 3, then, I present by-country estimates that include age in ranges (15–34, 35–59, 60–74, and 75+), factor variables for period, and cohorts aggregated into 5-year ranges. (Ekstam, 2021, shows that it is possible to use single birth-year variables for cohort. Taking that approach instead makes no difference to the results presented in Table 3.) The logic of the age ranges is intended to correspond to the substantive idea behind 'u-shapes': happiness should reach a low point in mid-life (35–59), rising as people get older (but then perhaps declining again as people become very old – this is the reason to distinguish between 60–74 and 75+). If age→happiness is genuinely u-shaped, then that relationship should appear in this approach as well. I extend this approach further below using a set of narrower ranges, with results presented in visual form.

**Table 3: OLS estimates of happiness, via age ranges**

|  | 15-34 | T | 60-74 | T | 75+ | T |
|---|---|---|---|---|---|---|
| Austria | 0.17 | 1.43 | 0.12 | 1.23 | 0.13 | 0.70 |
| Belgium | -0.05 | 0.66 | 0.15 | 2.21 | 0.05 | 0.39 |
| Bulgaria | -0.10 | 0.46 | -0.02 | 0.11 | -0.43 | 1.61 |
| Switzerland | 0.10 | 1.12 | 0.14 | 2.16 | 0.26 | 2.22 |
| Cyprus | -0.01 | 0.05 | -0.26 | 1.40 | -0.36 | 1.15 |
| Czechia | -0.01 | 0.10 | 0.24 | 3.04 | 0.16 | 1.01 |
| Germany | -0.05 | 0.62 | 0.13 | 2.02 | 0.12 | 1.10 |
| Denmark | 0.03 | 0.25 | 0.14 | 1.95 | 0.13 | 0.98 |
| Estonia | 0.01 | 0.09 | 0.08 | 0.84 | 0.10 | 0.71 |
| Spain | -0.03 | 0.43 | 0.15 | 1.91 | 0.21 | 1.70 |
| Finland | 0.00 | 0.04 | 0.18 | 3.31 | 0.10 | 1.03 |
| France | 0.03 | 0.37 | 0.17 | 2.27 | 0.32 | 2.47 |
| Britain | -0.15 | 1.65 | 0.34 | 4.63 | 0.30 | 2.37 |
| Greece | -0.04 | 0.21 | 0.02 | 0.15 | 0.16 | 0.70 |
| Croatia | -0.07 | 0.27 | -0.08 | 0.30 | 0.17 | 0.41 |
| Hungary | -0.01 | 0.10 | 0.30 | 3.09 | 0.28 | 1.65 |
| Ireland | -0.06 | 0.71 | 0.07 | 0.93 | 0.09 | 0.69 |
| Israel | 0.03 | 0.38 | 0.24 | 2.82 | 0.27 | 1.83 |
| Iceland | 0.04 | 0.24 | 0.14 | 0.87 | -0.07 | 0.23 |
| Italy | -0.13 | 0.78 | -0.04 | 0.28 | -0.47 | 1.87 |
| Lithuania | 0.10 | 0.64 | 0.15 | 0.97 | 0.22 | 0.90 |
| Luxembourg | 1.18 | 1.10 | -0.03 | 0.09 | 0.27 | 0.37 |
| Netherlands | -0.05 | 0.62 | 0.05 | 0.93 | -0.02 | 0.22 |
| Norway | 0.25 | 2.93 | 0.13 | 1.89 | 0.09 | 0.74 |
| Poland | 0.23 | 2.17 | 0.13 | 1.47 | 0.20 | 1.28 |
| Portugal | 0.23 | 2.22 | 0.09 | 1.17 | 0.32 | 2.43 |
| Russia | 0.21 | 1.81 | 0.18 | 1.56 | 0.71 | 3.50 |
| Sweden | 0.08 | 0.91 | 0.25 | 3.77 | 0.18 | 1.55 |
| Slovenia | -0.03 | 0.27 | 0.17 | 1.77 | 0.35 | 2.18 |
| Slovakia | 0.10 | 0.60 | 0.20 | 1.59 | -0.19 | 0.83 |
| Turkey | 0.19 | 0.40 | 0.17 | 0.41 | -1.06 | 1.53 |
| Ukraine | -0.59 | 3.22 | -0.14 | 1.02 | 0.18 | 0.81 |

Note: the reference category is people aged 35-59. Models include variables for period (survey year) and cohort (5-year ranges).

In this table, the age range 35-59 is the reference category. To discern support for a u-shape, then, we would need to see positive coefficients for the 15-34 range *and* the 60-74 range. In instances where that pattern is evident, we can conclude that happiness declines towards middle age and subsequently rises (in line with the idea behind the 'u-shape').

In Table 3, country names are again shaded where the patterns are consistent with the idea of u-shapes, here using a threshold of T=1 (a more generous standard than the 1.5 threshold used by Blanchflower). The conventional pattern is evident for only seven countries: Austria, Switzerland, Luxembourg, Norway, Poland, Portugal, and Russia. For the rest we do not see support for a u-shape in conclusions reached via consideration of statistical significance (the T statistic).

All analyses come with a mix of advantages and disadvantages. A key advantage for Table 3 is that the models effectively control for any cohort effects. A key (and indeed obvious) disadvantage is that the age ranges might be too large to capture important trends. I therefore now turn to an analysis that uses a set of narrower ranges (using a width of 10 years: 15–24, 25-34, etc.). With a larger number of narrow ranges, it becomes harder to draw conclusions via statistical significance. But statistical significance is overrated anyway (Cohen 1997, Engman 2013). So, in Figure 2 I present results in graphical form, so that we can discern visually whether u-shapes are evident, in part via consideration of the extent to which happiness declines and then rises.

It is impossible to see u-shapes as a universal pattern in Figure 2. (The groupings in that figure are not meaningful; they are mostly alphabetical, but certain exceptions were made to avoid crowding/overlap in the lines.) Inevitably, a visual judgement is subjective, to some extent; here we lack the (potentially misleading) simplicity of decision by asterisk ($p<0.05$). In subjective terms, then, u-shapes are reasonably evident for: Austria (but happiness initially rises), Switzerland, Denmark, Germany, Greece, Spain, Cyprus, Great Britain, France, Iceland (with however a notable decline after 75), Israel, Netherlands, Norway, Poland, Russia, Slovenia, Sweden, and Ukraine (really more of a W). It is particularly difficult to discern u-shapes for Bulgaria, Czech Republic, Estonia, Finland, Hungary, Ireland, Italy, Portugal, Slovakia, and Turkey. For Belgium, Hungary, and Lithuania, I am only willing to say: perhaps.

It is worth reiterating where the lines in these figures come from. They are from OLS models built around ranges of age, where the models contain controls for period and cohort. There are of course no other control variables, again because none are needed (as there are no antecedents of age). With controls for period and cohort, a cross-sectional comparison of people in one age range to people in another (older) range gives a reasonably good indication of what happens to people's happiness as they get older. (In the next section I offer a necessary caveat.) In general a cross-sectional comparison of this sort would be vulnerable to the possibility that people with different values of X are different in other ways that actually generate the patterns we see in an unadjusted relationship. But in this context, where X is age, there are no other variables (apart from period and cohort) that could generate that relationship, because there are no antecedents of age. The patterns in the figures, then, are reasonably interpreted as the causal effects of ageing on people's happiness (again because they are net of period and cohort effects).

**Figure 2: Models of age → happiness using 8 age ranges**

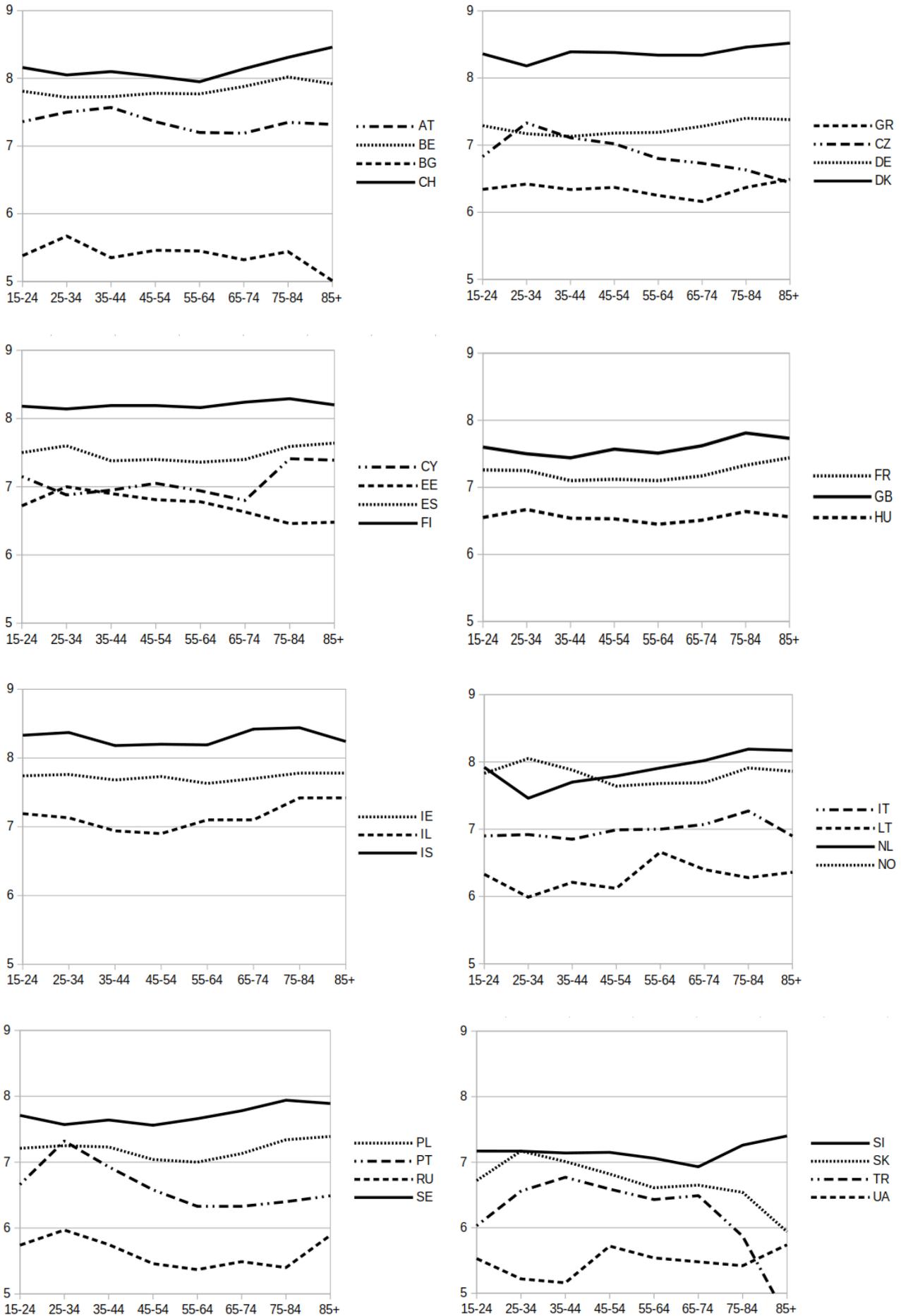

To offer a numerical consideration of effect size, in Table 4 I present the model-adjusted levels of life satisfaction in each age range by country (the numbers that give us the patterns in Figure 2). In the final three columns, we see the maximum and minimum for each country, followed by the difference. Where there are u-shapes, we can then gauge their magnitude. The largest differences are found for countries that most obviously do *not* have u-shapes (Turkey, Slovakia, Portugal, and Czech Republic). If we exclude these and also the other countries where (in my subjective judgement) u-shapes do not appear (Bulgaria, Estonia, Finland, Ireland, and Italy), we can then calculate an average: 0.44. This number gives us a generous indication of how 'deep' the u-shapes in Europe might be. It is generous because it is only the difference between the maximum and the minimum; it does not tell us that on average in Europe happiness falls by 0.44 points and then rises by 0.44 points – it only tells us that it does one or the other. In Blanchflower's analysis of ESS data (Figure 7), the depth of the u-shape curve for Europe is almost a full point on the 11-point scale – but here we see a number less than half of that.

[Table 4 is in landscape orientation and will not fit in the text; it is found at the end]

## *2.5 A deductive consideration of (hypothetical) longitudinal results*

The analyses in this paper are cross-sectional. (The same is true of the analyses in Blanchflower's 2021 article.) Use of cross-sectional data is inevitable in many instances; not all countries have panel datasets (and in some instances the panel data that are available do not contain questions on happiness). When cross-sectional data are used, it is then important to consider angles that the analysis cannot address.

In research comparing cross-sectional and longitudinal results (e.g. Frijters and Beatton 2012, Kratz and Brüderl 2021), the introduction of fixed effects (when using panel data) leads to results that indicate a (steeper) decline in happiness as people move through old age. Use of fixed effects corrects for a potential sample selectivity pattern, i.e., the likelihood that panel datasets oversample happier older people (and naturally omit older people whose unhappiness might have contributed to earlier mortality). Frijters and Beatton (2012) also suggest that panel datasets might undersample happier middle-age people; the samples would then give more weight to middle-age people who are relatively unhappy. In both respects an analysis that does not correct for these selectivity biases would increase the impression of a u-shape, especially by indicating a post-middle-age increase in happiness that is not 'real'.

With cross-sectional data, it is of course impossible to correct for bias rooted in these patterns. Investigations that work in a cross-sectional mode, then, must reflect on their results in these terms. The cross-sectional results presented above likely overstate any upward change in happiness as people move through old age; the real change is likely to be lower (i.e., more negative). For countries where results above appear to support a 'u-shape' finding (because happiness appears to rise after middle-age), a note of caution is therefore required: this pattern might disappear in a longitudinal analysis. Correction for selectivity (if possible, via use of fixed effects) would likely reveal a decline in happiness in older age (or, at least, a less-pronounced increase).

## 3 Conclusion

A finding of 'u-shapes everywhere' is not robust to reasonable alternative methodological decisions – at least not for European countries. An analysis using different methodological decisions produces results that for some countries depart (sometimes substantially) from the idea of u-shapes.

There are good reasons to prefer those methodological decisions over the decisions that produced the results in Blanchflower 2021. To estimate the age→happiness effect, we are best served by an analysis that meets the following criteria:

1. Use of the full range of adult ages; if happiness tends to decline in very old age (when people are likely to experience significant challenges), then a restriction on the upper bound will inflate the coefficients in a quadratic age function.
2. No controls for individual circumstances, where X→W; if controls of that sort are included, the result is likewise inflation of coefficients in a quadratic function.
3. Inclusion of controls for cohort and period, so that coefficients for age variables do not reflect underlying change for all and/or the formative experiences associated with when people were born.
4. Evaluation of results using a specified functional form via comparison to less restrictive analyses (in part so that conclusions take effect size into account and are not unduly driven by hypothesis testing, sample size, and $p<0.05$).

In general, a substantive question is usefully explored not just via one particular analytical approach but via several. Ideally, the results would be consistent across the different approaches (in the mode of 'robustness checks'). Here the results are not consistent in that way. So, we have to consider the relative merits of different approaches.

The analysis above that meets the four criteria most effectively comes with the results presented via Figure 2 (and Table 4). The quadratic coefficients in Table 2 impose a restrictive functional form (giving too much weight to hypothesis tests and $p<0.05$) and do not include a control for cohort. The less restrictive results presented in Table 3 (where a control for cohort is used) use a set of crude age ranges. Insofar as those results show u-shapes in roughly one-quarter of the countries analysed, however, there is no good reason to dismiss those findings. The potential problem with crude age ranges is that they might fail to detect u-shapes in *other* countries.

That possibility is realised only to a limited extent in the results presented in Figure 2. Those results are of course not without their own limitations. They do not include a consideration of statistical significance, which might be considered by some as a limitation (though it might also constitute a strength). If it is a limitation, then it is reasonably judged a less important limitation than the problems identified for the other approaches (the use of control variables where X→W, in particular, which leads to substantial bias). A reasonable summary of the findings presented above tells us that there is reasonably consistent evidence for a u-shape in 16 countries: Austria, Switzerland, Germany, Denmark, Spain, France, Britain, Israel, Iceland, Netherlands, Norway, Poland, Russia, Sweden, Slovenia, and Ukraine. (Again, however, a longitudinal analysis might undermine that conclusion for at least some of these countries.) For nine of the 30, the evidence weighs *against* a u-shape (Bulgaria, Czech Republic, Ireland, Italy, Portugal, Slovakia, Turkey). For the remaining 5 the various methods used here do not lead to a consistent finding.

I therefore conclude that we do not see u-shapes 'everywhere' in Europe. Figure 2 indicates a variety of patterns, some of them looking nothing like a u-shape. Tables 3 and 4 offer a reasonable indication that a u-shape describes the age→happiness relationship in *some* European countries, though by no means for all (or even most). Where u-shapes exist, they are also substantially 'shallower' than is portrayed in Blanchflower (2021). Whether the alternative methodological decisions used here would lead to a similar story in other world regions (and for other dependent variables) is left to future work. A similar outcome does not seem unlikely.

The debate about age→happiness u-shapes has raged for an extended period. That debate will no doubt continue. Ideally, it would continue not simply via presentation of a new set of results but via consideration of why one set of results differs from another. A resolution can come (if at all) only via that sort of comparative evaluation. The key strength of this article is not just the findings per se but the use of detailed methodological arguments for preferring these findings over others.

**Table 4: Levels of happiness by country and age range (adjusted for period and cohort effects)**

|  | 15-24 | 25-34 | 35-44 | 45-54 | 55-64 | 65-74 | 75-84 | 85+ | max | min | difference |
|---|---|---|---|---|---|---|---|---|---|---|---|
| Austria | 7.36 | 7.50 | 7.57 | 7.36 | 7.20 | 7.19 | 7.35 | 7.32 | 7.57 | 7.19 | 0.38 |
| Belgium | 7.81 | 7.72 | 7.73 | 7.78 | 7.77 | 7.88 | 8.02 | 7.92 | 8.02 | 7.72 | 0.30 |
| Bulgaria | 5.38 | 5.67 | 5.35 | 5.46 | 5.45 | 5.32 | 5.44 | 5.01 | 5.67 | 5.01 | 0.66 |
| Switzerland | 8.16 | 8.05 | 8.10 | 8.03 | 7.95 | 8.14 | 8.31 | 8.46 | 8.46 | 7.95 | 0.51 |
| Cyprus | 7.15 | 6.88 | 6.95 | 7.05 | 6.94 | 6.80 | 7.41 | 7.39 | 7.41 | 6.80 | 0.61 |
| Czech Rep. | 6.83 | 7.33 | 7.11 | 7.02 | 6.80 | 6.73 | 6.63 | 6.44 | 7.33 | 6.44 | 0.89 |
| Germany | 7.29 | 7.17 | 7.13 | 7.18 | 7.19 | 7.28 | 7.40 | 7.38 | 7.40 | 7.13 | 0.27 |
| Denmark | 8.36 | 8.18 | 8.39 | 8.38 | 8.34 | 8.34 | 8.46 | 8.52 | 8.52 | 8.18 | 0.34 |
| Estonia | 6.72 | 7.00 | 6.90 | 6.81 | 6.78 | 6.63 | 6.46 | 6.48 | 7.00 | 6.46 | 0.54 |
| Spain | 7.50 | 7.60 | 7.38 | 7.40 | 7.36 | 7.40 | 7.59 | 7.64 | 7.64 | 7.36 | 0.28 |
| Finland | 8.18 | 8.14 | 8.19 | 8.19 | 8.16 | 8.24 | 8.29 | 8.20 | 8.29 | 8.14 | 0.15 |
| France | 7.26 | 7.25 | 7.10 | 7.12 | 7.10 | 7.17 | 7.33 | 7.44 | 7.44 | 7.10 | 0.34 |
| Britain | 7.60 | 7.50 | 7.44 | 7.57 | 7.51 | 7.62 | 7.81 | 7.73 | 7.81 | 7.44 | 0.37 |
| Greece | 6.34 | 6.42 | 6.34 | 6.37 | 6.25 | 6.16 | 6.37 | 6.49 | 6.49 | 6.16 | 0.33 |
| Hungary | 6.55 | 6.67 | 6.54 | 6.53 | 6.45 | 6.51 | 6.64 | 6.56 | 6.67 | 6.45 | 0.22 |
| Ireland | 7.74 | 7.76 | 7.68 | 7.73 | 7.63 | 7.70 | 7.78 | 7.78 | 7.78 | 7.63 | 0.15 |
| Israel | 7.19 | 7.13 | 6.94 | 6.90 | 7.10 | 7.10 | 7.42 | 7.42 | 7.42 | 6.90 | 0.52 |
| Iceland | 8.33 | 8.37 | 8.18 | 8.20 | 8.19 | 8.42 | 8.44 | 8.24 | 8.44 | 8.18 | 0.26 |
| Italy | 6.90 | 6.92 | 6.85 | 6.99 | 7.00 | 7.07 | 7.27 | 6.90 | 7.27 | 6.85 | 0.42 |
| Lithuania | 6.33 | 5.99 | 6.21 | 6.12 | 6.66 | 6.40 | 6.28 | 6.36 | 6.66 | 5.99 | 0.67 |
| Netherlands | 7.92 | 7.46 | 7.70 | 7.79 | 7.91 | 8.02 | 8.19 | 8.17 | 8.19 | 7.46 | 0.73 |
| Norway | 7.83 | 8.05 | 7.88 | 7.64 | 7.68 | 7.69 | 7.91 | 7.86 | 8.05 | 7.64 | 0.41 |
| Poland | 7.21 | 7.25 | 7.23 | 7.04 | 7.00 | 7.13 | 7.34 | 7.39 | 7.39 | 7.00 | 0.39 |
| Portugal | 6.66 | 7.32 | 6.93 | 6.58 | 6.33 | 6.33 | 6.40 | 6.49 | 7.32 | 6.33 | 0.99 |
| Russia | 5.74 | 5.97 | 5.75 | 5.46 | 5.37 | 5.49 | 5.40 | 5.89 | 5.97 | 5.37 | 0.60 |
| Sweden | 7.71 | 7.57 | 7.64 | 7.56 | 7.66 | 7.78 | 7.94 | 7.89 | 7.94 | 7.56 | 0.38 |
| Slovenia | 7.17 | 7.17 | 7.14 | 7.15 | 7.06 | 6.93 | 7.26 | 7.40 | 7.40 | 6.93 | 0.47 |
| Slovakia | 6.72 | 7.17 | 7.01 | 6.82 | 6.61 | 6.65 | 6.54 | 5.94 | 7.17 | 5.94 | 1.23 |
| Turkey | 6.03 | 6.56 | 6.77 | 6.59 | 6.43 | 6.49 | 5.87 | 4.63 | 6.77 | 4.63 | 2.14 |
| Ukraine | 5.53 | 5.22 | 5.16 | 5.72 | 5.54 | 5.48 | 5.42 | 5.74 | 5.74 | 5.16 | 0.58 |